\documentclass[aip,cha,reprint,numerical]{revtex4-1}
\pdfoutput=1
\usepackage{graphicx}   
\usepackage[skip=1pt,font={sf,footnotesize}]{caption}
\usepackage{color}      
\usepackage{subcaption}

\usepackage{hyperref}  
\usepackage{amsmath}
\numberwithin{equation}{section}
\usepackage{amssymb}
\usepackage{verbatim}
\usepackage{float}
\usepackage{natbib}

\usepackage{chngcntr}
\counterwithout{equation}{section}

\usepackage{array} 
\usepackage{tabularx}
\usepackage{booktabs}
\usepackage[utf8]{inputenc}
\usepackage{bm}

\begin{document}
\title{Repulsively Coupled Kuramoto-Sakaguchi Phase Oscillators Ensemble Subject to Common Noise}
\author{Chen Chris Gong}
\email[]{cgong@uni-potsdam.de}

\affiliation{Institute of Physics and Astronomy, University of Potsdam, Karl-Liebknecht-Stra\ss e 32, 14476 Potsdam, Germany}

\author{Chunming Zheng}
\email[]{chzheng@uni-potsdam.de}
\affiliation{Institute of Physics and Astronomy, University of Potsdam, Karl-Liebknecht-Stra\ss e 32, 14476 Potsdam, Germany}

\author{Ralf Toenjes}
\email[]{ralf.toenjes@uni-potsdam.de}
\affiliation{Institute of Physics and Astronomy, University of Potsdam, Karl-Liebknecht-Stra\ss e 32, 14476 Potsdam, Germany}

\author{Arkady Pikovsky}
\email[]{pikovsky@uni-potsdam.de}
\affiliation{Institute of Physics and Astronomy, University of Potsdam, Karl-Liebknecht-Stra\ss e 32, 14476 Potsdam, Germany}
\affiliation{Department of Control Theory, Nizhny Novgorod State University,
Gagarin Avenue 23, 606950 Nizhny Novgorod, Russia}

\date{\today}

\begin{abstract}
We consider the Kuramoto-Sakaguchi model of identical
coupled phase oscillators with a common noisy forcing.
While common noise always tends to synchronize the oscillators, a strong repulsive coupling prevents 
the fully synchronous state and leads to a nontrivial distribution of oscillator phases. In previous numerical simulations,
a formation of stable multicluster states has been observed in this regime. However we argue here, that
because identical phase oscillators in the Kuramoto-Sakaguchi model form a partially
integrable system according to the Watanabe-Strogatz theory, the formation of clusters is impossible.
Integrating with various time steps reveals that clustering is a numerical artifact, 
explained by the existence of higher order Fourier terms in the errors of the employed 
numerical integration schemes. Monitoring the induced change in certain 
integrals of motion we quantify these errors. We support these
observations by  showing, on the basis of the analysis of the corresponding Fokker-Planck equation, 
that two-cluster states are non-attractive.  On the other hand, in ensembles of 
general limit cycle oscillators, such as Van der Pol oscillators, due to an 
anharmonic phase response function, as well as additional amplitude dynamics, 
multiclusters can occur naturally.
\end{abstract}


\maketitle

\begin{quotation}
Coupled oscillators can synchronize, if the nature of coupling makes their phases attract each other (attractive coupling);
or desynchronize, if there is repulsion of the phases (repulsive, or inhibitory coupling). This effect is well captured by the Kuramoto model of coupled
phase oscillators. Another way to synchronize oscillators is to act on them with a common force -- in particular,
even common random noise will bring the phases together, an effect known as noise-induced synchronization. 
An interplay of repulsive coupling with common noise can be nontrivial, since
the two effects act against each other. We argue in this paper that the Kuramoto model in such a case
cannot form multiple clusters, due to its special integrability based on the Watanabe-Strogatz theory. The only possible end state for an evolution under Kuramoto-Sakaguchi model with repulsive coupling and weak noise is complete asynchrony of the oscillators.
However, clusters can be observed in numerical simulations, because a discretization of the dynamics
in general breaks the integrability. We also show that in a more realistic model like
coupled Van der Pol oscillators, clusters can typically form due to the anharmonic phase response intrinsic to these oscillators.
\end{quotation}

\section{Introduction} \label{intro}
The Kuramoto model, since its formulation in 1975 by Y.Kuramoto \cite{Kuramoto75}, has been vastly successful
across scientific disciplines in describing naturally occurring phenomena in coupled oscillatory systems.
On the one hand its simple mathematical form allows the analytical solution of a mean field theory for coupled 
oscillators with non-identical natural frequencies in the infinite system size limit \cite{Kuramoto75,kura86}. On the other 
hand, it is still able to generate complex behavior, e.g. chimeras \cite{chimera1,chimera2}, chaos \cite{kurachaos} etc. 
The significance of the Kuramoto model to complexity science is akin to that of a simple model organism in the study of genetics, as its 
simple mathematical form nevertheless provides qualitative and quantitative insight into a variety of complex phenomena, especially as a paradigm 
for synchronization.

Since the original publications, there has been a plethora of literature dedicated to modelling systems by coupled Kuramoto oscillators.
A particularly idealised and simple model is that of the globally coupled identical oscillators. For such a system a powerful Watanabe-Strogatz (WS) theory \cite{ws94} 
exists, according to which identical phase oscillators under identical forcing in their frequencies and first harmonics evolve under a Lie group 
belonging to the general class of M{\"o}bius group action \cite{2009Chaos..19d3104M}. 
Such systems possess a hidden low-dimensional dynamics. Moreover,
it implies partial integrability, as it guarantees the existence of
constants of motion, which are preserved throughout the dynamics \cite{ws94}. The system 
is only ``partially'' integrable
because there are still 3 variables that are not constant and follow
nontrivial dynamical equations. Closely related to the WS approach 
is the Ott-Antonsen theory \cite{OA2018,OA2019}, which allows an exact formulation of the mean field dynamics 
(with 2 degrees of freedom) for an infinite-sized population of non-identical oscillators.

Beyond the discovery of periodic components of a complex dynamics, 
the theoretical significance of the WS integrability is still not quite clear. For example, the integrability
cannot be stated with certainty to exist in synchronous states, because in this case a cluster is in complete synchrony
 the same way as if the synchronized system had been ``non-integrable''. However, in situations
with partial synchrony, including the evolution towards aforementioned fully synchronized state,
the WS integrability leads to a multiplicity
of regimes due to the integrals of motion, and appear to have
additional periodic components in the dynamics (cf. application
of WS integrability to chimera states in Ref.~\onlinecite{PRzZ2}).
Recently, a perturbation theory based on the WS integrability
has been suggested~\cite{Vlasov-Rosenblum-Pikovsky-16}, which allows one to follow the evolution
of the integrals under perturbations. One of the important consequences
of the WS approach, which we explore below, is that it leads to the restriction on possible not
fully synchronous states -- namely, it excludes the formation of several clusters.

This study is inspired by the results of Gil et al.~\cite{Mikhailov}, where clustered states
 were observed in a set of identical oscillators
under common multiplicative noise and repulsive coupling. In general, globally coupled
identical oscillators can demonstrate configurations of different structural complexity: complete synchrony
(one cluster state), partial synchrony (a nontrivial continuous distribution of phases,
where all individual phases can be different), clusters 
(several groups of fully synchronized oscillators), chimeras 
(when one or several cluster coexist with partially synchronous oscillators), 
and solitary states~\cite{solitarystate} (when only one oscillator with a different phase exists apart 
from the fully synchronous cluster). Under strong repulsive coupling, a
fully synchronized cluster becomes unstable, and it is not
evident a priori which of these aforementioned configurations will be observed. Therefore, Gil et al.~\cite{Mikhailov}
conducted numerical simulations and reported that common noise generally causes clustering in globally repulsively 
coupled interactions. 

We call this claim into question based on the following reasons. First,
clustering is indeed observed in some models of globally coupled identical phase
oscillators~\cite{cluster1, cluster2, syncbook}, but always in situations
 with complex
interaction functions, i.e, when the coupling term includes higher harmonics of the coupled phases. 
However, no such higher harmonics were
present in the interaction term in the model proposed by Gil et al. 
Secondly, recent studies of
the competition between common noise and
repulsive coupling revealed non-trivial distributions for identical and non-identical 
oscillators \cite{Pimenova16, Mason}, but no clustering has been observed. 
Gil et al., on the other hand, reported that for
 identical oscillators, clusters formed without a threshold, at any noise intensity. 
 Finally, because the model 
used by Gil et al. can be fully described by the WS theory, there are restrictions due to 
the general properties of the M\"obius transform governing the dynamics \cite{2009Chaos..19d3104M}, 
i.e. clusters are not allowed to appear (see Section~\ref{sec:wst} below). 
Therefore, a thorough numerical and analytical study is needed to resolve
the conflict between numerical findings~\cite{Mikhailov} and known theories.

 In this paper we will 
thoroughly study the formation and stability of clusters in numerical experiments. After formulation of the problem 
(section~\ref{sec:Prob}) we will present the WS approach and show that clustering is impossible. This conclusion
will be further supported by an analytic and numerical investigation of the Lyapunov exponents of oscillators evolving 
in the field of two fully synchronized clusters (section \ref{sec:linstab}).
In terms of numerics, the exact integrability is not preserved by the
standard numerical schemes, both for deterministic and stochastic equations. In section~\ref{sec:num} we analyse errors in 
different numerical methods in terms of the change in the constants of motion which should be preserved by the M\"obius group action, and show that this 
can lead to the formation of clusters as a numerical artifact.
The WS approach is restricted to oscillators which couple to common external forces in their first harmonics. Furthermore, the
Langevin equations must be understood in the Stratonovich interpretation in order for the usual differential calculus used in the WS approach to be applicable.
Since we attribute the formation of clusters to the violation of WS integrability, we test this hypothesis in section~\ref{sec:noise}
by including higher order terms in the phase dynamics, and in section \ref{sec:vdp}
by studying repulsively coupled Van der Pol oscillators under common additive noise, for which higher order Fourier terms and multiplicative noise are naturally present
in the phase reduced dynamics.
\section{Problem Formulation} \label{sec:Prob}
We study a population of identical phase oscillators with phases $\{\varphi_k\}$ subject
to a global coupling of Kuramoto-Sakaguchi type \cite{kura86} and common phase-dependent noise terms
\begin{equation} \label{eq:kurasaka2noise}
\begin{aligned}
\dot{ \varphi}_{k} = &\omega + \frac{1}{N}\sum\limits_{j = 1}^{N}\sin(\varphi_{j}- \varphi_{k} + \gamma) \\
&+\sigma_1 \eta_1(t) \sin\varphi_{k} + \sigma_2 \eta_2(t) \cos\varphi_{k}~.
\end{aligned}
\end{equation}
Here $\eta_{1,2}(t)$ are Gaussian white noise terms, with $\langle \eta_{i}(t) \rangle=0$, 
$\langle \eta_i(t)\eta_j(t') \rangle = \delta_{ij}\delta(t-t')$~. 

The parameters $\sigma_1$ and $\sigma_2$ parametrize the noise 
strengths for the two noise terms. 
The Langevin Eq.\eqref{eq:kurasaka2noise} is to be interpreted in the Stratonovich sense so that WS theory is applicable.
Indeed, when interpreted in the sense of It\^o calculus, for $\sigma_1\ne\sigma_2$ a noise-induced drift, i.e. Stratonovich shift, exists, and is proportional to the second harmonics in $\varphi_k$ which violates the conditions of WS theory.
The phase shift parameter $\gamma$ 
parametrizes the degree of repulsion and attraction in the coupling term. 
In particular, when $\gamma = 0$, the coupling is purely attractive, for $\gamma = \pi$, it is purely repulsive, 
and for $\gamma = \pi/2$ the coupling is neutral. Because it is always possible to rescale time, we 
assume, without loss of generality, that the phases couple with unit strength to the mean field in Eq.\eqref{eq:kurasaka2noise}.

Models of type \eqref{eq:kurasaka2noise} have been analysed in Ref. \onlinecite{Nagai-Kori-10} 
for the case of one noise term, and in Gil et al. \cite{Mikhailov} for two noise terms.
In the latter work it has been argued, that model \eqref{eq:kurasaka2noise} is the proper 
approximation after phase reduction,
for a population of weakly coupled identical Stuart-Landau oscillators with a 
common additive noise term which is isotropic in the complex plane, making the phase equations invariant under rotation. In this case, $\sigma_1=\sigma_2=\sigma$
and one can rewrite \eqref{eq:kurasaka2noise} as
\begin{equation} \label{eq:saka2noise2}
\dot{\varphi}_{k}=\omega+\mathrm{Im}[(Ze^{{i} \gamma}+\sigma \xi)e^{-{i}\varphi_{k}}]~, 
\end{equation}
where $Z = \frac{1}{N}\sum\limits_{j=1}^N{e^{{i}\varphi_j}}$ is the Kuramoto 
mean field and $\xi=-\eta_1+i\eta_2$ is complex Gaussian white noise.
There exist well-known results on the stability of the completely synchronous 
cluster $\varphi_1=\varphi_2=\ldots=\varphi_N$
for model \eqref{eq:saka2noise2}:
To quantify the degree of stability of a fully synchronous cluster, 
which corresponds to $|Z| = 1$,  one calculates the transversal 
Lyapunov exponent (in previous literature also known as ``evaporation'' or ``split Lyapunov exponent''~\cite{Pimenova16}), which describes
the evolution of oscillator phases slightly deviated from the cluster. 
It is their average exponential rate of approach towards a cluster (or the rate of moving away from the cluster if the exponent is positive). The expression for
this exponent is ~\cite{Pimenova16}
\[
\lambda=-\cos\gamma-\frac{\sigma^2}{2} ~.
\]
For a negative Lyapunov exponent, complete synchronization is stable, i.e. it 
attracts nearby phases that are perturbed from it.
According to this formula, for strong enough noise, the cluster 
of complete synchrony, with $|Z|=1$, is stable. 
For repulsive coupling, with $\cos\gamma<0$ and weak enough
noise, the cluster is unstable. 

While in Ref.~\onlinecite{Pimenova16} mainly the statistical properties of $|Z|$ have been analysed, 
Ref.~\onlinecite{Mikhailov} focuses on the occurrence of clusters, which are distinct attractive subgroups of oscillators with 
identical phases within the groups. The main goal of
this paper is twofold: (i) to demonstrate that the occurrence of clusters 
in system~\eqref{eq:saka2noise2} is impossible,
and (ii) to identify numerical artefacts that may nevertheless lead to cluster 
formation in simulations.

\section{Theory} \label{sec:theory}
\subsection{Application  of the Watanabe-Strogatz theory} \label{sec:wst}
Our basic equation \eqref{eq:saka2noise2} in the Stratonovich interpretation belongs to a class of problems for 
which a theory, first developed by Watanabe and Strogatz~ \cite{ws94} in 1994, which we 
shall call WS theory, is applicable. WS theory reduces the  N-dimensional dynamics of a system of identically driven 
identical phase oscillators
\begin{equation}
\dot\varphi_k=\omega(t)+\textrm{Im}[H(t) e^{-{i} \varphi_k}],\quad k=1,\ldots,N
\label{eq:gws}
\end{equation}
where $\omega(t)$ and $H(t)$ are arbitrary real and complex-valued functions of time, respectively,
to a three-dimensional dynamical system preserving $N-3$ independent constants of motion. It is evident, 
that Eq.~\eqref{eq:saka2noise2} belongs to class~\eqref{eq:gws}. 
One must stress here, that qualitative arguments below are applicable to any common
force acting on the oscillators, not necessarily to the white Gaussian
noise case mainly treated in this paper.
It can be colored noise, or a chaotic/quasi-periodic/periodic force, or any combination of these
functions of time.

At the heart of WS theory (see Refs.~\onlinecite{ws94,PR08} for a detailed presentation) 
is a coordinate transformation $\mathcal{M}$ formally belonging to the class of M{\"o}bius 
maps, which is a type of fractional linear transformation, mapping the complex unit circle one-to-one to itself. Specifically, $\mathcal{M}$ and its inverse $\mathcal{M}^{-1}$ 
can be written as
\begin{align} \label{eq:moebtrafo}
\mathcal{M}&: \psi_{k} \rightarrow \varphi_{k}, \hspace{3mm} e^{{i} \varphi_{k}}  = \frac{ z+e^{{i}(\psi_{k}+\beta)} }{ 1+z^{*} e^{{i}(\psi_{k}+\beta)} } ~, \\
\mathcal{M}^{-1}&: \varphi_{k} \rightarrow \psi_{k}, \hspace{3mm}  e^{i \psi_{k}} = e^{-i\beta}\frac{ z - e^{{i}\varphi_{k}} }{ z^{*} e^{{i}\varphi_{k}} - 1} ~.
\label{eq:invmt}
\end{align}
Here $\{\varphi_{k}\}$ are the original phase coordinates, $z$ is a complex parameter 
of absolute value smaller than 1 and the parameter $\beta$ is a rotation angle.
If the $\varphi_k$ evolve according to \eqref{eq:gws} and $z$ and $\beta$ evolve according to
\begin{align} \label{eq:WS}
\dot z &= {i}\omega(t) z + \frac{1}{2} H(t) - \frac{1}{2} H^{*}(t) z^{2},  \\
\dot \beta &= \omega(t) + \textrm{Im}[z^{*} H(t)] ~,\nonumber
\end{align}
then the $\psi_k=\mathcal{M}^{-1}(\varphi_k)$ are time independent constants of motion. The constants of motion are determined by the actual phases $\varphi_k$ and three time-dependent, real-valued parameters, the amplitude and angle of $z$ and by $\beta$. 
It is possible to impose conditions on the constants which make the M{\"o}bius transform unique. For instance, if no majority cluster exists one can choose $z$ and $\beta$ such that $\langle \exp(i\psi_k) \rangle = 0$ and $\textrm{arg}\left(\langle\exp(i2\psi_k)\rangle\right)=0$ [\onlinecite{ws94}].
The M{\"o}bius transform \eqref{eq:moebtrafo} consists essentially of a common rotation of the angles $\psi_n$ by the angle $\beta$ and a subsequent contraction along the circle into the direction of the angle of $z$. Indeed, $|z|$ can be used as a measure of synchronization akin to the Kuramoto order parameter as both become equal to unity at complete synchronization.
The existence of the constants of motion implies the system is integrable. However, it must be stressed that 
WS integrability only holds for phase oscillator models of the form Eq.~\eqref{eq:gws}, e.g. the Kuramoto-Sakaguchi 
model for globally coupled phase oscillators or the Winfree model \cite{winfree} with a harmonic phase response function. 

Equations \eqref{eq:WS} in principle allow for a numerical integration of the system
which automatically conserves all the constants $\psi_k$. However, typically the forcing term $H(t)$ contains the Kuramoto order parameter $Z=\langle\exp(i\varphi_k)\rangle_k$ for which the $\varphi_k=\mathcal{M}(\psi_k)$ must be calculated at every time step. Only for constants of motion which are uniformly distributed
on the unit circle the parameter $z$ approximates the actual 
mean field $Z$ to the order $1/\sqrt{N}$ where $N$ is the number of oscillators \cite{PR08}\cite{PRzZ2}.

Expressions \eqref{eq:moebtrafo} and \eqref{eq:invmt} alone already rule out 
the formation of clustered states from non-clustered initial conditions. Two oscillators with initially different phases are mapped to a single point only if $|z|\to 1$, in which case $z_k=\exp({i}\varphi_k)$ tends to $ z$ for all points of the circle except a singular ``solitary state'' point with $z+\exp\left(i(\psi_k+\beta)\right)=0$ [\onlinecite{solitarystate}] for which the M\"obius transform is not unique at $|z|\to 1$. Therefore, only a single cluster attracting different phases can exist at a time. Nevertheless, this only prohibits the formation of multiple clusters but not their existence under this model. There is not restriction for oscillators to be in one or several distinct clusters with identical
phases within each cluster and stay in that configuration. 

\subsection{Linear stability analysis of a two-cluster state} \label{sec:linstab}
In section \ref{sec:wst} above we have shown that clusters cannot appear from non-clustered 
initial conditions. The same arguments can be applied when the dynamics evolves from an initial multicluster state.
Here, either the multicluster remains with the same partition, 
or the fully synchronized state with maximally one
additional cluster or oscillator in a solitary state appears. Imperfect clusters, i.e.
configurations with phases very close to one another, can also dissolve or contract depending on their dynamical stability.
It is therefore instructive to look on the stability of the multicluster states.  According to 
the WS theory, one expects that not more than one of the clusters can be asymptotically attracting.
Otherwise multiclusters would also form from non-clustered initial conditions, a phenomenon which is forbidden by the argument in section \ref{sec:wst}. In this 
section we provide a linear stability analysis of the two-cluster state 
under the stochastic evolution given by model \eqref{eq:saka2noise2} which confirms our expectation.
We write equations~\eqref{eq:saka2noise2} for a two-cluster state as
\begin{equation}\begin{aligned} \label{eqn:2-clus-dyn}
\dot{\Phi}_{1}=&\omega + p_{1}\sin\gamma+p_{2}\sin(\Delta\Phi+\gamma)\\&+\sigma\sin\Phi_{1}\eta_1(t)+\sigma\cos\Phi_{1}\eta_2(t) \\
\dot{\Phi}_{2}=&\omega + p_{2}\sin\gamma-p_{1}\sin(\Delta\Phi-\gamma)\\&+\sigma\sin\Phi_{2}\eta_1(t)+\sigma\cos\Phi_{2}\eta_2(t) ~.
\end{aligned}\end{equation}
Here $\Phi_1$ and $\Phi_2$ are the phases of the two clusters. $\Delta\Phi=\Phi_2-\Phi_1$ is their phase difference. 
Parameters $p_1$ and $p_2=1-p_1$ are their relative population sizes.

To evaluate the stability of the two-cluster state in terms of a small 
perturbation from one of the clusters, say, cluster 1, we perturb
two oscillators belonging to cluster 1 by pulling them by a small amount in opposite 
directions away from the cluster, i.e. $\varphi_{1,2} =\Phi_1\pm \delta$. For small $\delta$,
linearization yields
\begin{equation} \label{eq:2-clus-delta}
\begin{aligned}
\dot\delta=&\delta \big[-p_{1}\cos\gamma-p_{2}\cos(\Delta\Phi+\gamma)\\&
+\sigma\eta_1(t)\cos\Phi_{1}-\sigma\eta_2(t)\sin\Phi_{1}\big] ~.
\end{aligned}
\end{equation}
This allows us to express stability of cluster 1 via the split/evaporation 
Lyapunov exponent \cite{splitLE_Kaneko-89,splitLE_Pikovsky2001b,splitLE_pikovsky_politi_2016} as
\begin{equation} \label{eq:LE_ks}
\begin{aligned}
\lambda_1=&-p_{1}\cos\gamma-p_{2}\langle\cos(\Delta\Phi+\gamma)\rangle\\
&+\sigma\langle\eta_1(t)\cos\Phi_{1}\rangle-\sigma 
\langle\eta_2(t)\sin\Phi_{1}\rangle ~,
\end{aligned}
\end{equation} 
where $\langle \cdot \rangle$ indicates time average, which in this case also equals 
to the ensemble average, because as we will see in \eqref{eq:LE0}, the probability 
distribution of the phase $\Delta\Phi$ is stationary. 

\begin{figure}[ht]
\includegraphics[width=\columnwidth]{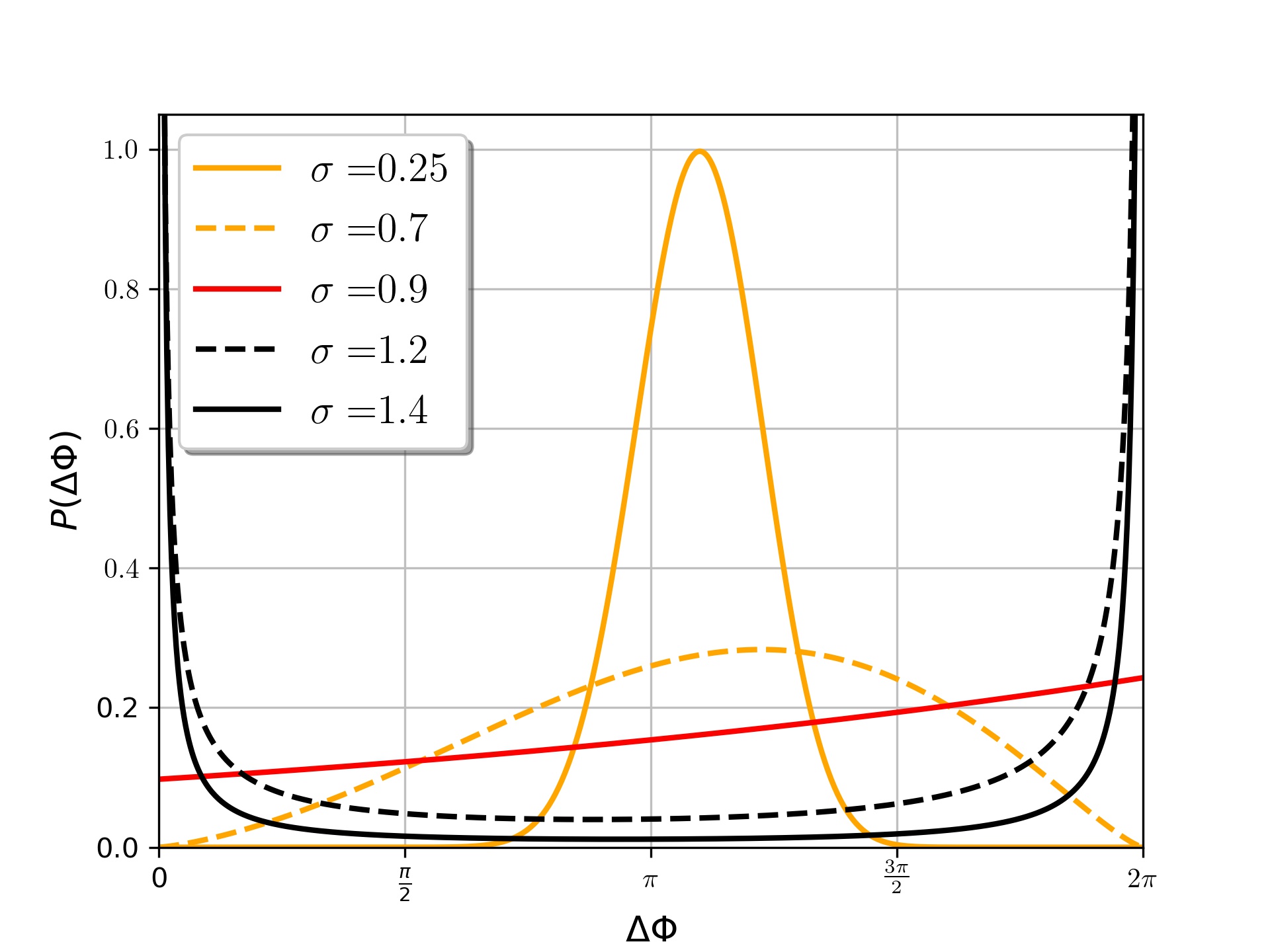}
	\caption{Probability density function \eqref{eq:LE0}, when $p_1 = 0.4$, $\gamma = 0.8\pi$ and the noise strength takes on various values (see legend). When the exponent of $\left|\sin\frac{\Delta\Phi}{2}\right|$ is positive, the function peaks asymmetrically at values larger than $\pi$ (orange lines, solid and dashed), and when the exponent is 0, it is an exponential distribution (red), and when it's negative, the distribution has an asymmetrical singularity at 0 (black lines, dashed and solid). The function will become a delta function when the exponent is $-1$ (not shown here).
	}
	\label{fig:probDelPhi}
\end{figure}

While the Stratonovich shift for the Langevin Eqs.\eqref{eqn:2-clus-dyn} happens to be zero, that is not the case anymore when Eq.\eqref{eq:2-clus-delta} is considered as well. It is important to keep this in mind and choose a correct integration scheme when Eqs.\eqref{eqn:2-clus-dyn} and \eqref{eq:2-clus-delta} are integrated numerically.
To calculate the Lyapunov exponent analytically, we need to know the probability distribution of $\Delta\Phi$
and the averages $\langle\eta_1(t)\cos\Phi_{1}\rangle,\;\langle\eta_2(t)\sin\Phi_{1}\rangle$.
First, we write a two-dimensional Fokker-Planck  equation for $\Phi_1$ and $\Delta\Phi$ corresponding
to the Langevin equations \eqref{eqn:2-clus-dyn} under Stratonovich interpretation \cite{HANGGI1982207}. 
Then, integrating the joint density of $\Phi_1$ and $\Delta\Phi$ over the
variable $\Phi_1$, using the fact that the probability distribution 
of $\Phi_1$ is rotational symmetric, i.e. uniform, we obtain a closed  
equation for the probability distribution $P(\Delta\Phi)$ of the phase difference
\begin{equation}
\begin{aligned}
&\frac{\partial P(\Delta\Phi)}{\partial t}=-\frac{\partial}{\partial\Delta\Phi}
\left\{\left[(p_{2}-p_{1})\sin\gamma(1-\cos\Delta\Phi)\right.\right.\\
&-\left.\left.\cos\gamma\sin\Delta\Phi\right]P\right\}+
\sigma^2\frac{\partial^{2}}{\partial \Delta\Phi^{2}}\left[(1-\cos \Delta\Phi)P\right] ~.
\end{aligned}
\end{equation}
Note that this probability density function is defined and restricted on the open interval $(0, 2\pi)$ since the two clusters cannot cross each other. The stationary solution has the form
\begin{equation}\label{eq:LE0}
P(\Delta\Phi)\sim \exp\left[\frac{\Delta\Phi(p_{2}-p_{1})\sin\gamma}{\sigma^2}\right]
\left|\sin\frac{\Delta\Phi}{2}\right|^{-2\left(\frac{\cos\gamma}{\sigma^2}+1\right)}.
\end{equation}
A closed expression for the normalized probability density is possible  when $\gamma=\pi$, i.e. the repulsion
between the oscillators is maximal. In this case,
\begin{align} \label{eq:LE1}
P(\Delta\Phi)&=\frac{1}{2B\left(\frac{1}{\sigma^2}-\frac{1}{2},\frac{1}{2}\right)}
\left|\sin\frac{\Delta\Phi}{2}\right|^{-2\left(1-\frac{1}{\sigma^2}\right)} ~,
\end{align}
where $B(x,y)$ is the Beta function. The shape of the probability density function in the general case \eqref{eq:LE0} is shown in Fig. \ref{fig:probDelPhi}.

In the expression \eqref{eq:LE0}, a nonzero phase shift parameter $\gamma$ introduces a curious asymmetry in form of the exponential factor which is not a periodic function, i.e. when we consider the distribution on $(0,2\pi)$ wrapped around the circle, the derivative is not continuous at the singular absorbing point $\Delta\Phi=0=2\pi$. 
The critical noise strength $\sigma_{c}$ beyond which the two clusters are synchronized 
to become one cluster, corresponds to the case where $P(\Delta\Phi)$ becomes a 
delta function $\delta(\Delta\Phi)$. Formally, this corresponds to divergence of the 
integral of the probability density~\eqref{eq:LE0}. 
This happens if the exponent of $|\sin(\Delta\Phi/2)|$ is smaller than $-1$, 
and from this we can calculate the critical noise strength to be $\sigma_c^2/2 = -\cos\gamma$.

In addition to the distribution of the phase difference $\Delta \Phi$, one needs 
to calculate the averages
$\langle\eta_{1}(t)\cos\Phi_{1}\rangle$ and $\langle\eta_{2}(t)\sin\Phi_{1}\rangle$ 
to evaluate \eqref{eq:LE_ks}. Since $\eta_1(t)$ and $\eta_2(t)$ are independent Gaussian white noise processes and $\Phi_1[\eta_1,\eta_2]$ is a functional of both $\eta_1$ and $\eta_2$ this can be accomplished by virtue 
of the Furutsu-Novikov formula \cite{Furutsu63,Novikov64}
\begin{equation} \label{Eq:FN}
\begin{gathered}
\langle \eta_2(t) \sin\Phi_1 \rangle = \int \delta(t-t')\left\langle  \frac{\delta \sin\Phi_1}{\delta \eta_2} 
\right\rangle dt'= \\
\left\langle\frac{d (\sin\Phi_1)}{d \Phi_1}\frac{\delta \Phi_1}{\delta \eta_2}\right\rangle= \left\langle\frac{1}{2}\sigma \cos^{2}\Phi_1\right\rangle = \frac{\sigma}{4} ~.
\end{gathered}
\end{equation}
Similarly $\langle \eta_1(t) \cos\Phi_1 \rangle = -	\frac{\sigma}{4}$.
A general expression for the Lyapunov exponent $\lambda_{1}$, which describes the stability 
of the cluster 1, is therefore
\begin{equation}
\lambda_{1}=-p_{1}\cos\gamma-(1-p_{1})\int_{0}^{2\pi}\cos(\Delta\Phi+\gamma)P(\Delta\Phi)d\Delta\Phi-\frac{\sigma^2}{2}\;.
\label{eq:le}
\end{equation}
Lyapunov exponent~\eqref{eq:le} can even be analytically represented for the case $\gamma=\pi$
\begin{equation} \label{eq:special_lambda}
\lambda_{1}=\left\{  
\begin{array}{lr}  
p_{1}+p_{2}(\sigma^2-1)-\frac{\sigma^2}{2}, &\sigma^2<2\;; \\  
1-\frac{\sigma^2}{2}, &\sigma^2\geq 2    \;.
\end{array}  
\right.  
\end{equation}
Exchanging $p_1$ and $p_2$, we obtain the Lyapunov exponent $\lambda_2$ of the other cluster. 
From this special case one can easily see that when $\sigma^2<2$, i.e. when a fully synchronized one-cluster 
state is unstable and the two-cluster Lyapunov exponents are well defined,
they satisfy $\lambda_{1}+\lambda_{2}=0$. 

\begin{figure}[ht]
\includegraphics[width=\columnwidth]{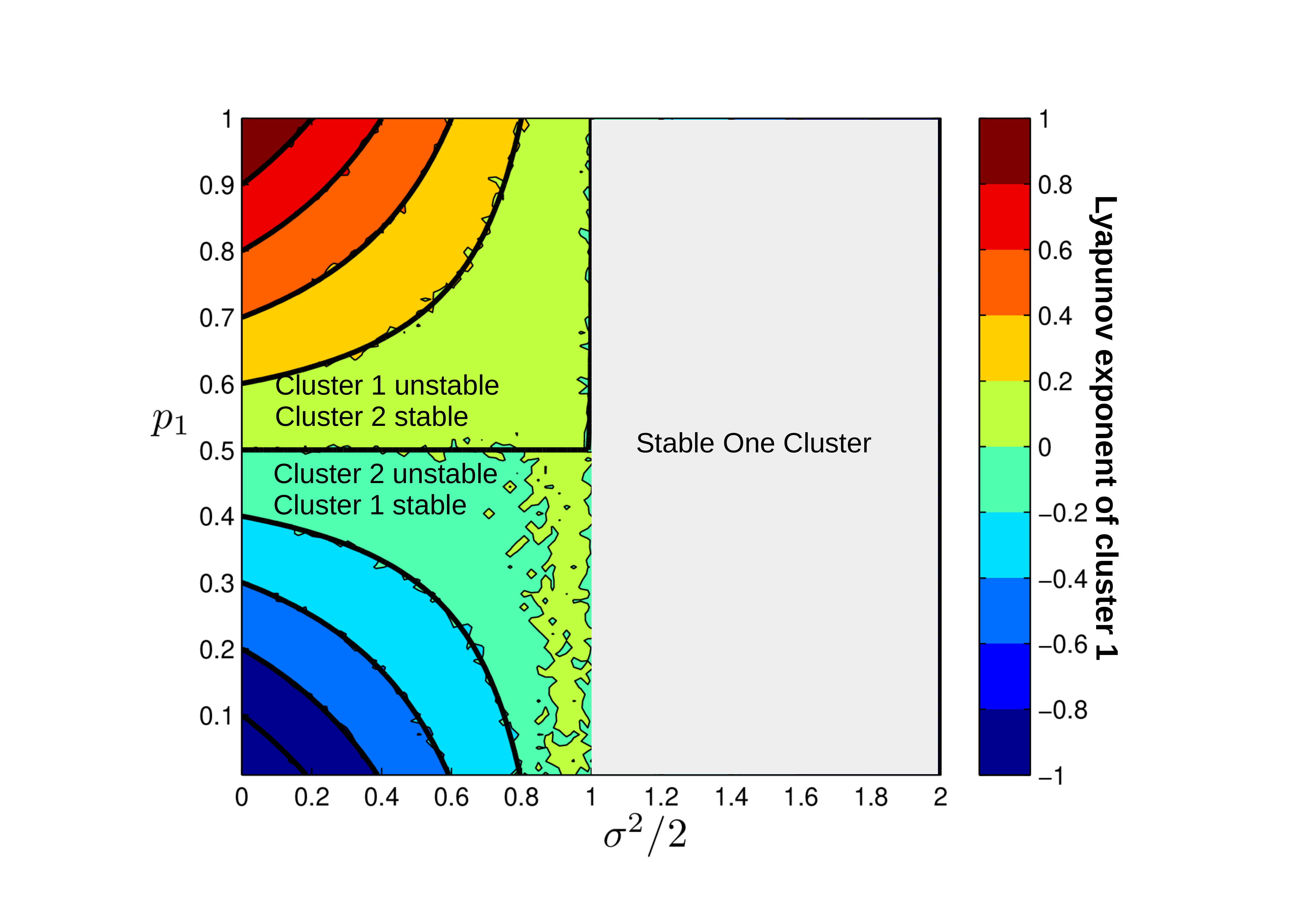}
	\caption{Diagram for linear stability of cluster 1 of the two clusters indicated by its Lyapunov exponent, for 
	phase shift $\gamma=\pi$ (maximal repulsion),
	in the plane of parameters $p_1$, the relative size of cluster 1, and the noise strength $\sigma^2/2$.
	Bold solid lines: theoretical result \eqref{eq:special_lambda} obtained 
	by Fokker-Planck formulation. Contour lines/color:  
	by direct simulation of Eqs. \eqref{eqn:2-clus-dyn}-\eqref{eq:LE_ks} via Euler-Heun scheme. The Lyapunov 
	exponent for cluster 1 below the critical noise strength $\sigma^2_{c}/2 = 1$ is shown in color gradient.
	Above the critical noise strength one cluster is formed. The diagram is symmetric with respect to the line 
	 $p_1=0.5$ (except for very small positive exponents for $p_1<0$ and $\sigma^2\approx 2$, 
	 which can be attributed to finite averaging time), indicating that together with the second cluster Lyapunov exponent $\lambda_1+\lambda_2= 0$.
	}
	\label{fig:LE}
\end{figure}

Through direct numerical evaluation of the Lyapunov exponent 
$\lambda_1$ in Fig.~\ref{fig:LE}~, we obtain a confirmation of the above analytical result. 
 
For the case of Kuramoto-Sakaguchi model with general phase shift parameter $\gamma$, 
using~\eqref{eq:LE_ks} and the corresponding expression for $\lambda_2$, we obtain for the sum
\begin{equation} 
\begin{split}
\lambda_1+\lambda_{2} &= - \cos\gamma - \sigma ^2 - \cos\gamma \int_{0}^{2\pi} \cos\Delta\Phi P(\Delta\Phi)d\Delta\Phi \\
&+ (p_2-p_1)\sin\gamma\int_{0}^{2\pi} \sin\Delta\Phi P(\Delta\Phi)d\Delta\Phi
\end{split}
\end{equation} 
After applying integration by parts for the product of three functions, 
$\langle \sin\Delta\Phi \rangle$ can be written in terms of 
$\langle \cos^2(\Delta\Phi/2) \rangle$ and $\langle \cos\Delta\Phi \rangle$. After simple algebra, the relation 
$\lambda_1+\lambda_{2} = 0$ for the generic Kuramoto-Sakaguchi model can be shown. This means that for two narrowly distributed groups of repulsively coupled oscillators with common multiplicative noise, the larger group will dissolve while the smaller group is attractive (Fig.~\ref{fig:LE}). Simultaneous attraction into two clusters is not possible.
\section{Numerics}
\label{sec:num}
As we have argued above, the WS theory of integrability prevents a multicluster state from ever occurring in model
\eqref{eq:saka2noise2}, and a linear stability analysis via Fokker-Planck formulation has confirmed it in the case of two clusters in the repulsive regime. But the observation of clustering in simulations by Gil et al. \cite{Mikhailov}
may be attributed to numerical artefacts, as one cannot expect WS integrability to be preserved
by standard numerical methods. In this section we explore how different numerical integration methods 
for integrating deterministic and stochastic equations affect WS integrability and clustering.  
First we discuss methods which quantify the errors occurred from a deviation from integrability and measure the degree of clustering.

\subsection{Numerical evaluation of the constants of motion} \label{sec:com}
As has been outlined in section~\ref{sec:wst}, the constants of motion of the system
can be determined via the M\"obius transformation \eqref{eq:moebtrafo} of the $N$ phases $\{\varphi_{j}\}$. 
In practice, one must first determine the complex variable $z = \rho e^{{i} \Phi}$ which characterises the 
transformation. Watanabe and Strogatz (see section 4.2 in Ref. \onlinecite{ws94}) have shown that this can 
be accomplished with the help of a potential function
\begin{equation}
\mathcal{U}(\rho, \Phi) = \frac{1}{N} \sum\limits_{j} \log \frac{1 - 2\rho \cos(\varphi_{j}-\Phi) + \rho^{2}}{1 -\rho^{2}}~. 
\label{eq:wspot}
\end{equation}
The proper value of $z$ corresponds to the minimum of this function 
with respect to its modulus $\rho$ and to its argument $\Phi$. The easiest way to determine the minimum is by integrating
\[
\dot\rho=-\mathcal{U}_{\rho}\;,\qquad \dot\Phi=-\mathcal{U}_{\Phi}\;,
\]
until the steady state is established. The angles $\psi_k+\beta$ can then be obtained 
with the M{\"o}bius transformation \eqref{eq:moebtrafo}. To avoid determination of the value of $\beta$, 
it is convenient to consider only the differences $\psi_j-\psi_1$, $j=2,\ldots,N$ as constants of motion.
The disadvantage of this method lies in the necessity of solving the minimization 
problem for the potential \eqref{eq:wspot}, which can be performed with finite accuracy 
only. There exists, however, another possibility to determine the constants of motion.

Marvel et al. (see Sec. V in Ref.~\onlinecite{2009Chaos..19d3104M}) have demonstrated that the 
cross ratio of four complex numbers on the unit circle
is a preserved quantity under the M{\"o}bius transformation. 
For any four phases $\varphi_k,\varphi_{k+1},\varphi_{k+2},\varphi_{k+3}$ 
(not necessarily ordered on the circle), the constant of motion $C_{k}$ 
is defined as
\begin{equation} \begin{aligned}
\label{eq:constantMMS3}
&C_{k} = \frac{S_{k,k+2}}{S_{k,k+3}} \cdot \frac{S_{k+1,k+3}}{S_{k+1,k+2}}~, \text{~where } S_{ij} = \sin \frac{\varphi_i- \varphi_j}{2} ~.
\end{aligned} 
\end{equation}

Our method of checking the conservation of these quantities is based on \eqref{eq:constantMMS3},
 but we find it appropriate to avoid calculating fractions, 
because as the phases synchronize, the denominators can be very small or vanish. 
Instead we calculate the errors of the form
\begin{gather*}
e_k=S_{k,k+2}(t)S_{k+1,k+3}(t)S_{k,k+3}(0)S_{k+1,k+2}(0) -\\ 
S_{k,k+3}(t) S_{k+1,k+2}(t) S_{k,k+2}(0) S_{k+1,k+3}(0) ~.
\end{gather*}

In summary, we test for integrability in numerical schemes by calculating the following errors containing changes in the 
conserved quantities under M\"obius action
\begin{align} 
\text{Err}_{\text{WS}}(t) = &\max_k(\sin|(\psi_k(t)-\psi_1(t)) - (\psi_k(0)-\psi_1(0))|), \nonumber \\
&\text{where } k = 2,...N~, \label{eq:errs1}\\
\text{Err}_{\text{MMS}}(t) = &\max_k(|e_k|),\text{ where } k=1,2,\ldots,N-3~. \label{eq:errs2}
\end{align} 

\subsection{Numerical evaluation of clustering}

\begin{figure}[ht]
\includegraphics[width=\columnwidth]{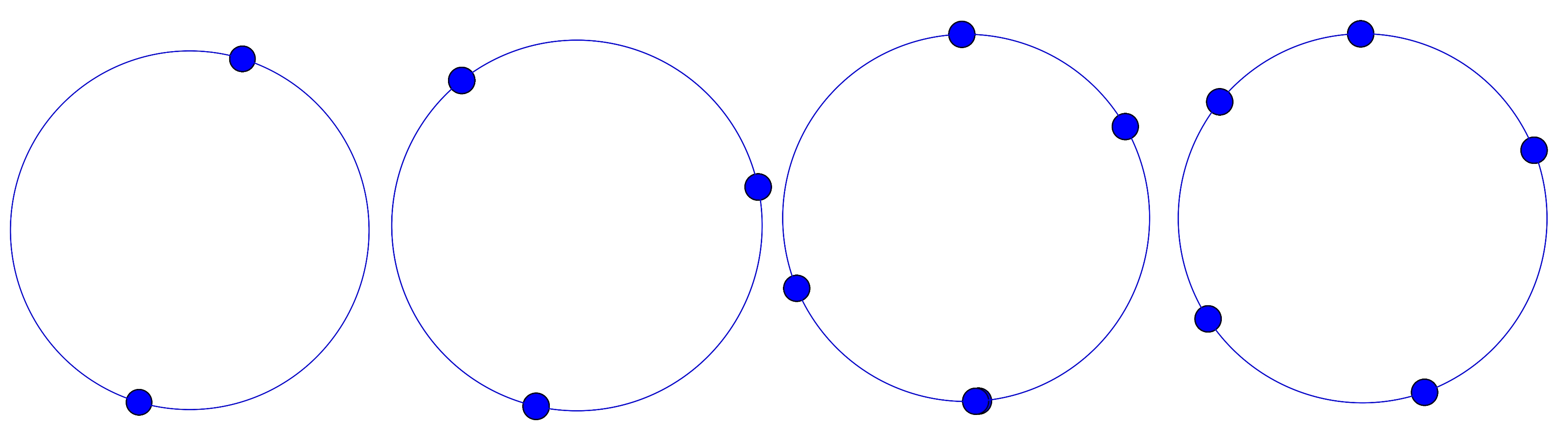} 
\caption{Besides 2-cluster and 3-cluster states discovered by Gil et. al \cite{Mikhailov}, 
4 or 5 clusters can also be found for large integration steps. Shown are 
multiclusters formed from the same initial condition of 100 Kuramoto 
phase oscillators, drawn randomly from a uniform distribution, with coupling phase 
shift $\gamma=2\pi\alpha \in (\pi/2, \pi)$ (repulsive coupling regime), various 
integration step sizes $h$ and noise strengths $\sigma$, after an 
integration time of $T$. (a): $\sigma=0.01$, $\alpha=0.3$, $h=1.0$, $T=220,000$; 
(b): $\sigma=0.1$, $\alpha=0.4$, $h=1.5$, $T=225,000$;   
(c): $\sigma=0.1$, $\alpha=0.4$, 
$h=2.0$, $T=100,000$; (d): $\sigma=0.1$, $\alpha=0.45$, $h=2.0$, $T=100,000$.
In most cases the final distributions of clusters are close to equipartition;
in some cases the dynamics is quite complex, with switchings between different nearly-clustered states.} \label{fig:kura}
\end{figure}

In numerical simulations of model \eqref{eq:saka2noise2} (details shall be outlined below)
we may observe different clustered states, illustrated in Fig.~\ref{fig:kura}.  

We quantify the formation of synchronized 
clusters with the help of the Kuramoto-Daido mean fields
\begin{align} \label{eq:Daido}
Z_k=\frac{1}{N}\sum_me^{{i} k \varphi_m} ~.
\end{align} 
After long integration time, the first order mean field $Z_1$ for repulsive coupling is 
either small if noise is present, or vanishes completely in the deterministic case.
The second order parameter $R_2=|Z_2|$ is maximal and equal to 1 for two fully synchronized clusters of arbitrary sizes with phase 
difference $\pi$ between them. Altogether, 
the degree of the formation of two clusters can be measured by a growth of $R_2$ 
approaching values close to one.
We will henceforth use the evolution of $R_2$ as an indication for a two-cluster state.

\subsection{Deterministic evolution}

\begin{figure}[t!]
\centering
\captionsetup[subfigure]{justification=centering}
	\begin{subfigure}[b]{0.75\columnwidth}
 \includegraphics[width=1\textwidth]{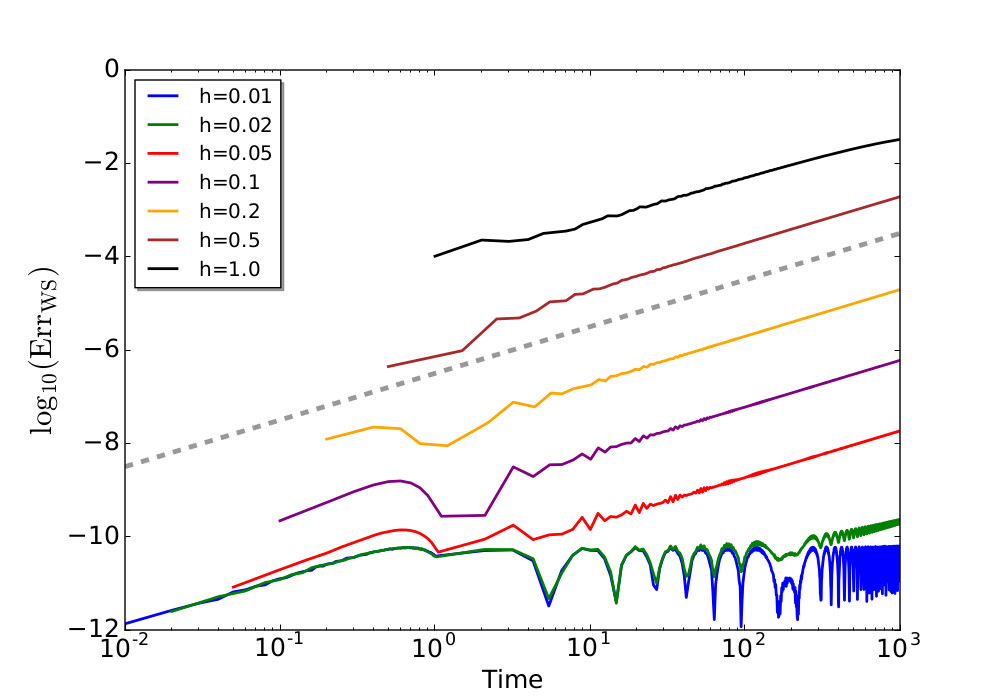} 
	 \subcaption{$\text{Err}_{\text{WS}}(t)$, RK4}
 	\end{subfigure}
	\begin{subfigure}[b]{0.75\columnwidth} 		
 \includegraphics[width=1\textwidth]{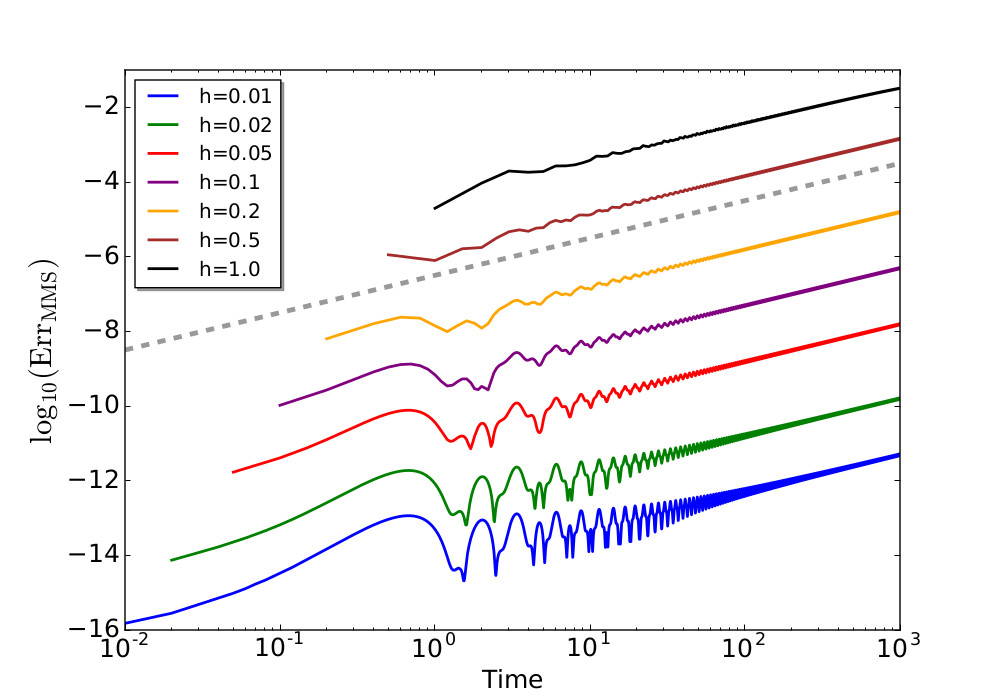}
	 \subcaption{$\text{Err}_{\text{MMS}}(t)$, RK4}
 	\end{subfigure} 
	\begin{subfigure}[b]{0.75\columnwidth} 		
 \includegraphics[width=1\textwidth]{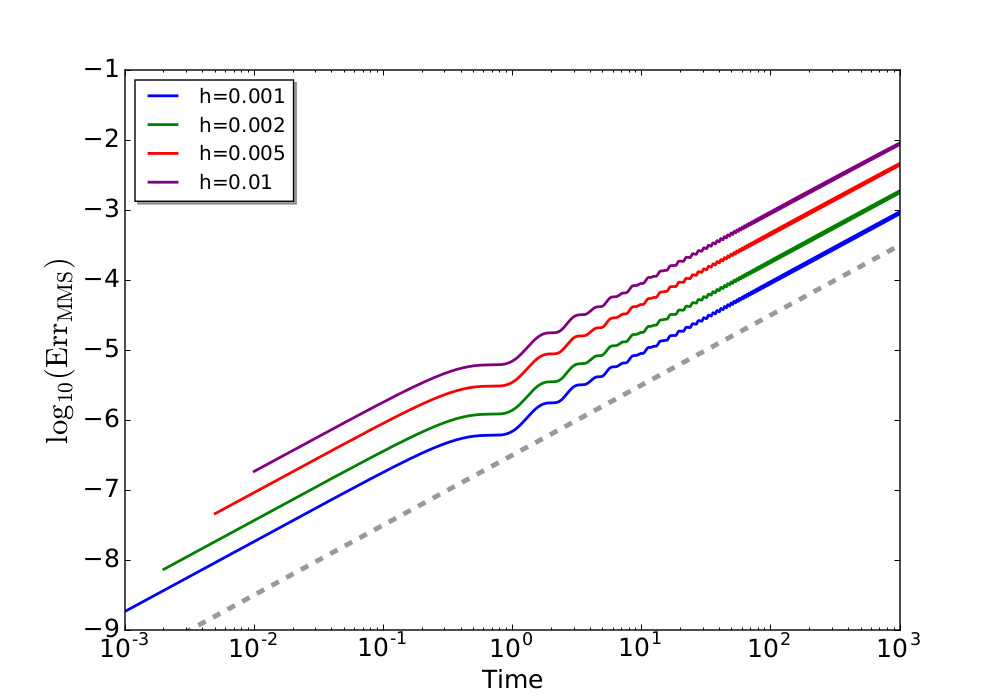}
	 \subcaption{$\text{Err}_{\text{MMS}}(t)$, Euler}
 	\end{subfigure}	
 	\begin{subfigure}[b]{0.75\columnwidth} 		
 \includegraphics[width=1\textwidth]{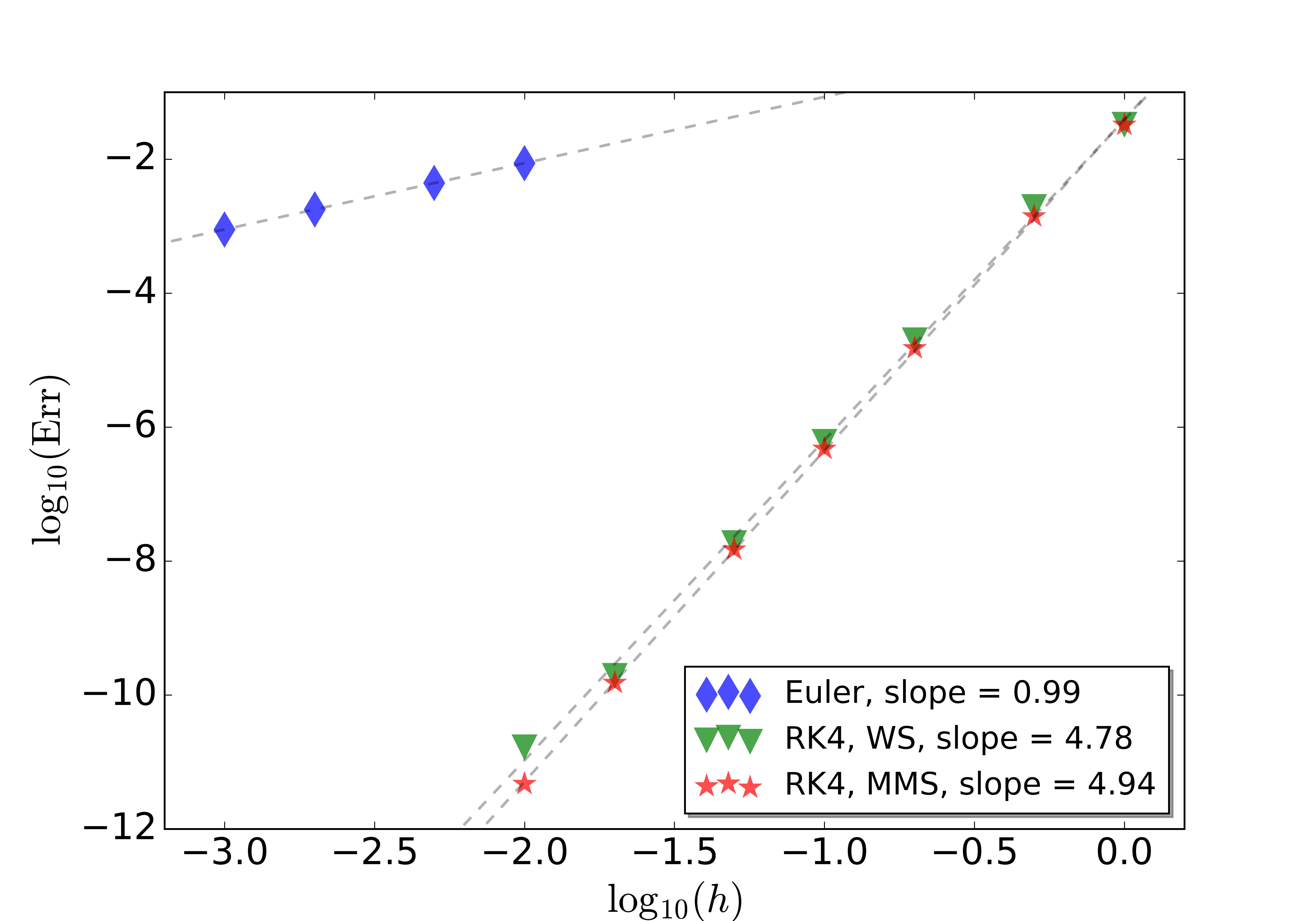}
	 \subcaption{$\log_{10}\text{Err}(T)$ vs. $\log_{10}h$}
 	\end{subfigure}	 	
  \caption{Time evolution of errors \eqref{eq:errs2} from integration of a deterministic equation with RK4 (a-b) and Euler (c) scheme. 
 Dashed lines in (a-c) have slope = 1 in log10-log10 plot. (d) shows the numerical errors at end time $T=1000$ vs. time step $h$, as well as their linear fits (shown in legend). \label{fig:const}}
\end{figure}

We first explore how well the WS integrability is preserved in numerical simulation of deterministic 
equations. Here the original Kuramoto-Sakaguchi model is not optimal. After a short 
initial transient, the evolution of $R_1=|Z_1|$ effectively comes to a halt as soon as a steady state is reached, i.e. $R_1=|Z_1|$ is zero for repulsive coupling or unity for attractive coupling. Instead, we integrate a model of type \eqref{eq:gws}
with a prescribed modulated time-dependent forcing $\omega (t) = 0.2\sin(1.752 t)$, 
$H(t)=0.4\cos(2.33 t) \cdot Z$, and $N=10$, designed to ensure the 
state remains nontrivial (see Fig.~\ref{fig:const}). Integrating this deterministic equation, we use the standard
Runge-Kutta method of 4th order (RK4) and the first-order Euler method. 

First, comparing Fig.~\ref{fig:const} panels (a) and (b), where the two methods \eqref{eq:errs1} and \eqref{eq:errs2}
of determining the constants of motion are used, we can conclude
that, while the errors in the constants of motion are similar for large steps,
 the calculation of the errors via $\text{Err}_{\text{WS}}$ \eqref{eq:errs1}  does 
 not allow for a proper estimation
of very small errors, due to the necessity of a minimization procedure which can be performed
only with finite precision. Therefore, for the rest of the paper we calculate 
the errors using only $\text{Err}_{\text{MMS}}$ \eqref{eq:errs2}.

The second observation is that in all the cases the errors grow in time roughly linear, 
with prefactors depending on the integration step $h$:
$\text{Err}_{\text{MMS,RK4}}\sim h^{4.94} t$ for the RK4 method, and $\text{Err}_{\text{Euler}}\sim h^{0.99} t$ 
for the Euler method, indicating a drift of the constants of motion. This is consistent with the fact that RK4 makes an error of $h^5$ in each time step and for Euler it is $h$.

\begin{figure}[ht]
 \includegraphics[width=\columnwidth]{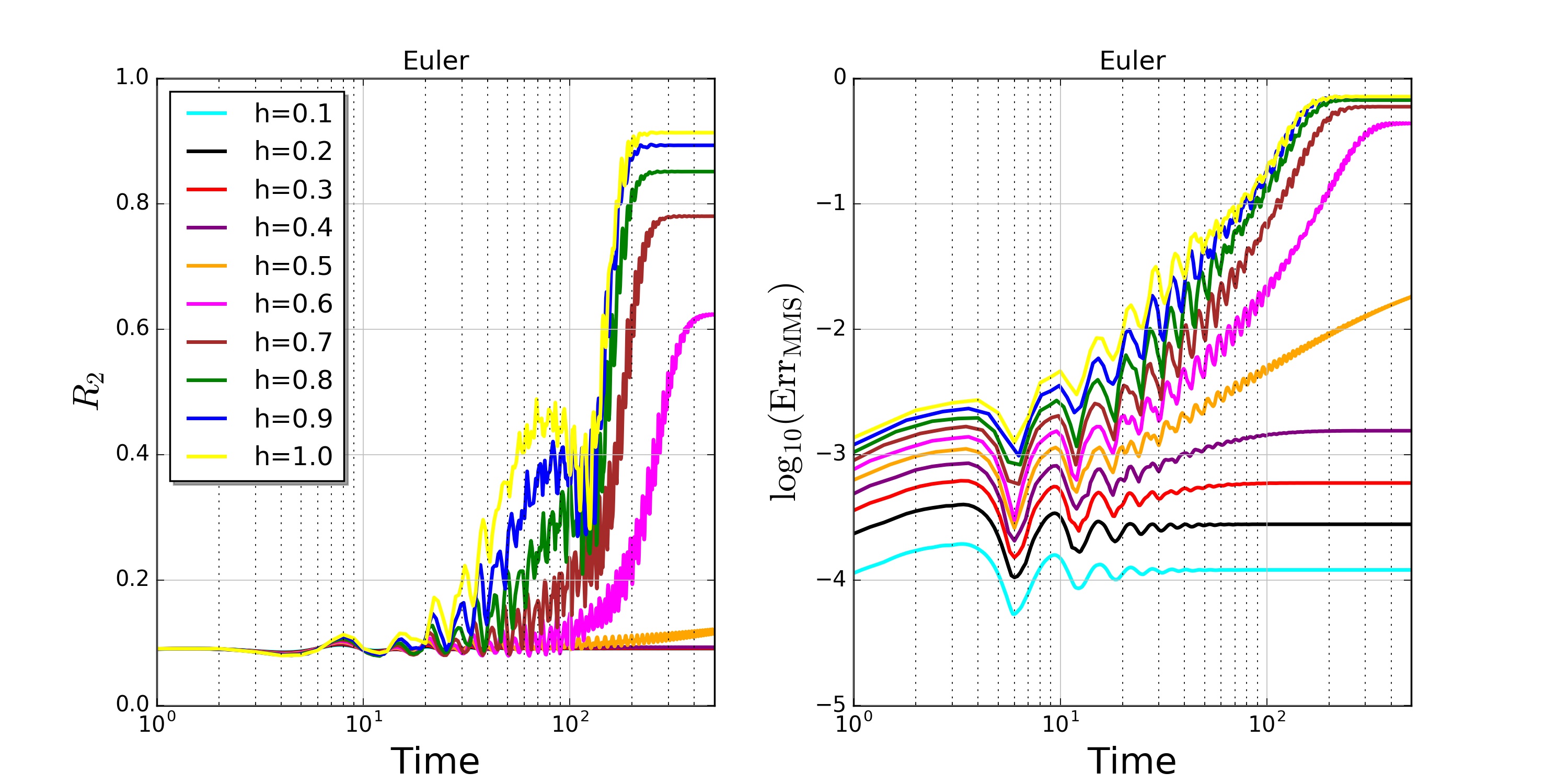}
 \caption{Second Kuramoto-Daido order parameter (left) \eqref{eq:Daido} and integrability errors (right) \eqref{eq:errs2} ,
 averaged over 10 random initial conditions drawn from uniform distribution,
as functions of time for Eq.~\eqref{eq:kurasaka2noise} 
 with $\omega=\sigma_1=\sigma_2=0$, $N=100$ and $\gamma=0.54\pi$. 
 Euler method is used for integration. Cluster formation on the left
 always corresponds to poor conservation of the constants on the right by the Euler scheme. 
}
 \label{fig:2panel0noise}
\end{figure}

In Fig.~\ref{fig:2panel0noise} we present the results for the 
integration of the deterministic equation \eqref{eq:kurasaka2noise} with 
$\omega=\sigma_1=\sigma_2=0$, $N=100$ and $\gamma=0.54\pi$ (slightly repulsive).
Here we use rather large integration steps to make the clustering effect visible 
during a relatively short transient time interval, before the main order parameter
becomes very small and the dynamics stops. 
One can see that for $h>0.6$ the order parameter $R_2$, which measures formation 
of a two-cluster state, grows to macroscopic values.

For instructive purposes, we explore here which type of perturbations are introduced 
by the numerical integration methods to the original dynamics \eqref{eq:kurasaka2noise} when noise is not present.
The simplest case is to estimate the perturbations introduced by the Euler method. The Euler method
models a continuous-time dynamical system $\dot\varphi_k=f_k(\vec{\varphi})$ up to the order $h$ as a map
$\varphi_k(t+h)=\varphi_k(t)+h f_k(\vec{\varphi}(t))$. Then we might ask, what 
continuous equation is integrated by the same map correctly up to the order $h^2$.
Looking for this equation in the form of $\dot\varphi_k= f_k(\vec{\varphi})+
h g_k(\vec{\varphi})$, we find
\[
g_k(\vec{\varphi})=-\frac{1}{2}\sum_m f_m(\vec{\varphi})
\frac{\partial}{\partial\varphi_m} f_k(\vec{\varphi}) ~.
\] 
Substituting for the Kuramoto-Sakaguchi model $f_k=\omega+\textrm{Im}\left[Ze^{{i}\left(\gamma-\varphi_k\right)}\right]$,
we obtain a modified equation where the error in the Euler integration is part of the dynamics
\begin{eqnarray}
\label{eq:KuramotoEulerError}
\dot \varphi_k &=&\omega+\textrm{Im}\left[Z e^{{i}\left(\gamma-\varphi_k\right)}\right] - \\&&
\frac{\epsilon}{4}\textrm{Im}\left[Z e^{{i}\left(2\gamma-\varphi_k\right)} - Z^* Z_{2}e^{-{i}\varphi_k} - Z^2 e^{{i} 2\left(\gamma-\varphi_k\right)}\right]~.\nonumber
\end{eqnarray}
Here $Z_2=\left\langle\exp\left(i2\varphi_m\right)\right\rangle_m$ is the second Kuramoto-Daido mean field. 
One can see, that in addition to the new
coupling terms proportional to $\sin\varphi$ or $\cos\varphi$, which do not violate the WS integrability, terms
proportional to $\sin2\varphi,\;\cos2\varphi$ appear, which violate the WS integrability and may result
in the formation of two clusters. Clusters of higher orders can presumably be attributed to terms
$\sim \sin 3\varphi$ etc. appearing in higher order errors in $h$.

\subsection{Stochastic evolution} \label{sec:noise}
\begin{figure}[ht]
 \includegraphics[width=\columnwidth]{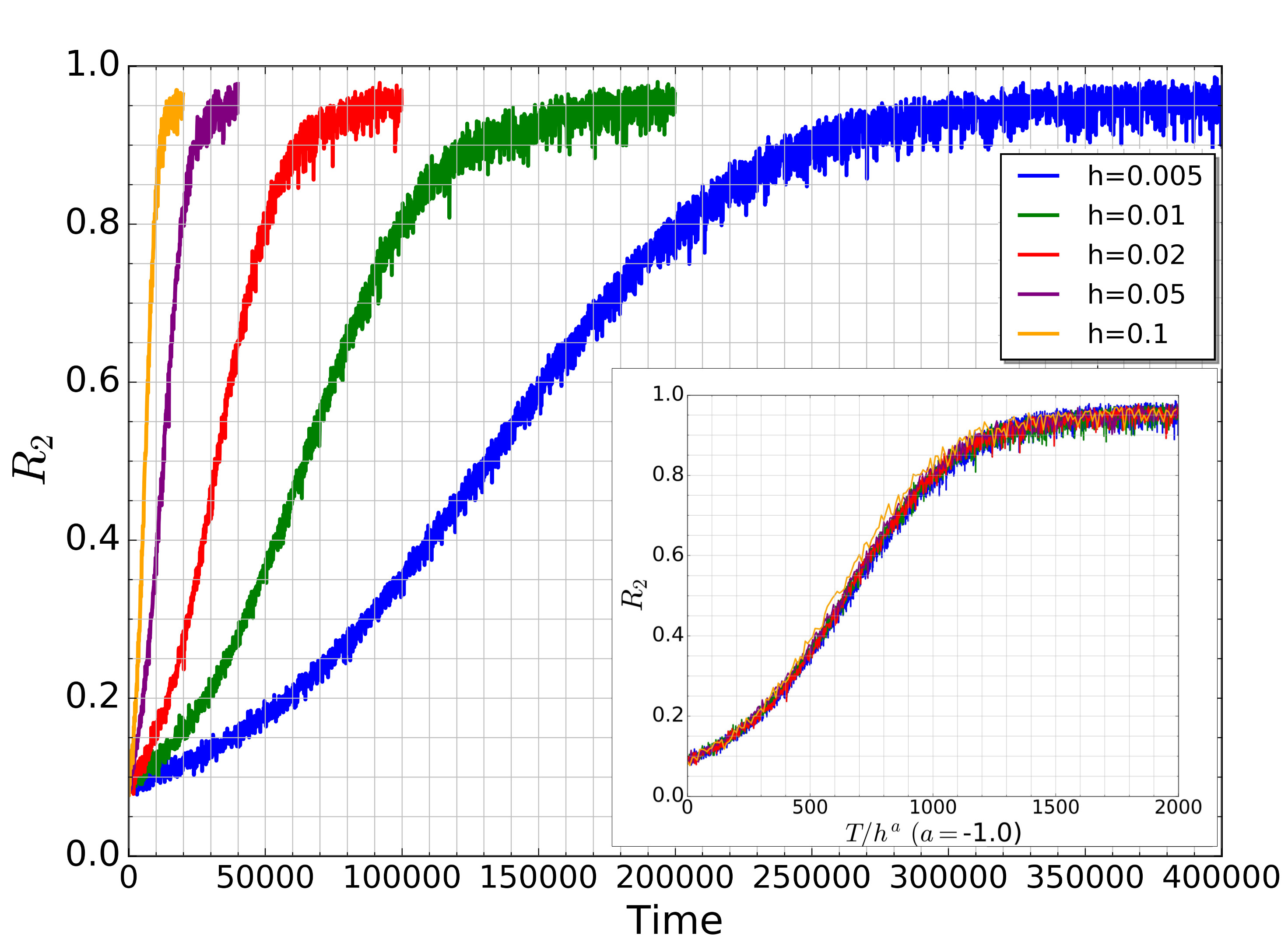}
 \caption{Evolution of the second Kuramoto-Daido order parameter $R_2$  under the Euler-Heun 
 scheme for different step sizes integrating \eqref{eq:kurasaka2noise} with strong noise $\sigma_1=\sigma_2= 0.1$. Inset: collapse of the curves when plotted as functions of  $h t$. 
 The data is averaged over simulations 
 with 10 different sets of initial conditions. System parameters: system size 
 $N=100$, intrinsic frequency $\omega = 0$, noise strength $\sigma = 0.1$, phase shift $\gamma=0.6\pi$.}\label{fig:R2}
\end{figure}

\begin{figure}[ht]
 \includegraphics[width=\columnwidth]{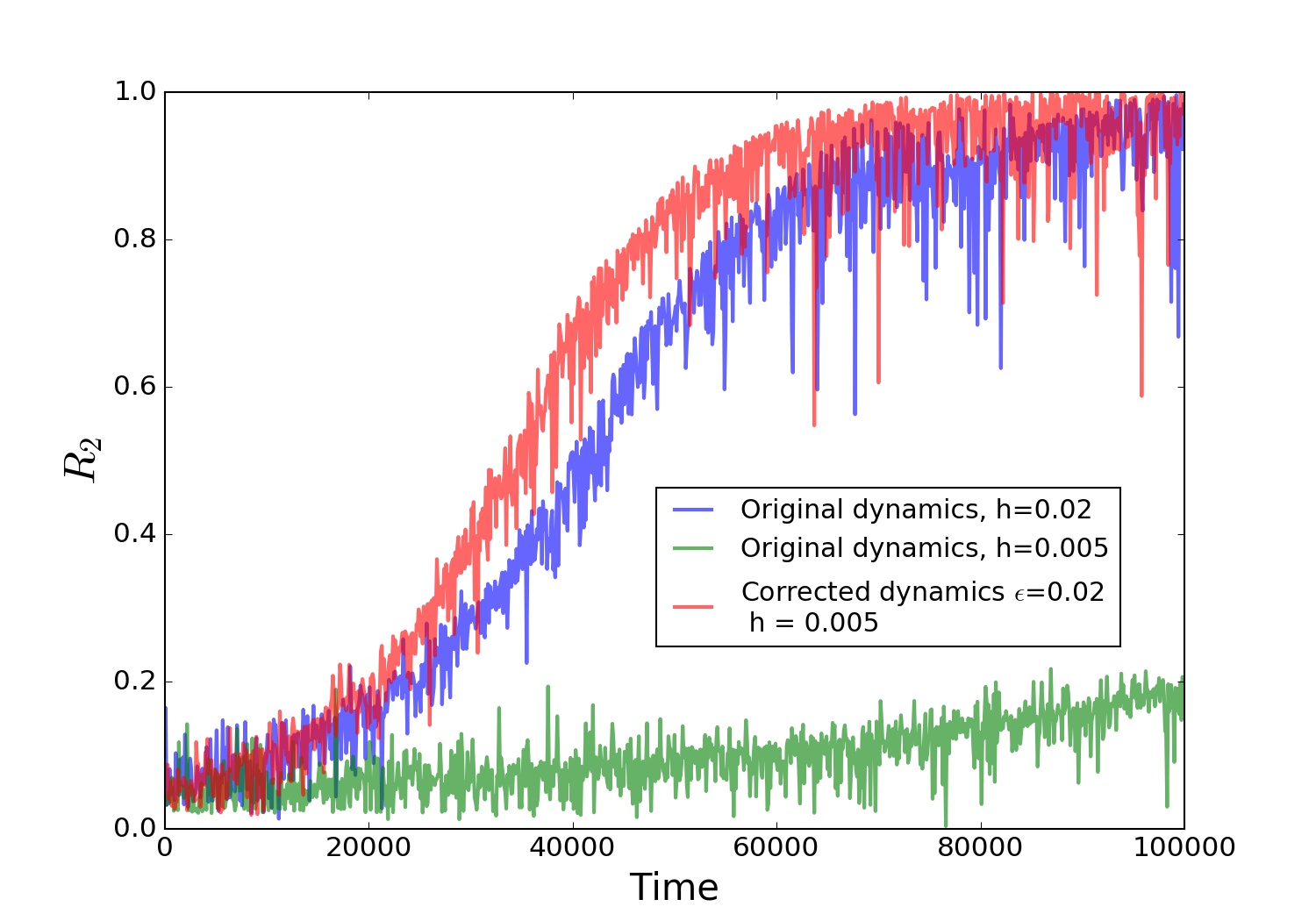}
 \caption{Three Euler-Heun integrations of identical initial conditions showing the time series for the 2-cluster order parameter $R_2$. Blue and green: under the original model Eq.~\eqref{eq:kurasaka2noise} with 2 equal noise terms, with time step $h_0 = 0.02$ and $h_1 = 0.005$, respectively. Red: under the modified dynamics Eq.~\eqref{eq:KuramotoEulerError} with the perturbation amplitude $\epsilon=h_0$ in the modified dynamics and time steps $h=h_1$. The time scales at which the 2 clusters build up for blue and red time series are comparable, supporting the hypothesis that the Fourier terms of second order in the discretization errors of the integration scheme are responsible for the formation of two clusters.}\label{fig:2timescales}
\end{figure}

Throughout this paper we interpret the stochastic system \eqref{eq:kurasaka2noise}
in the Stratonovich sense. However, the additional drift term needed to 
transform it into It\^{o} interpretation, vanishes for the case when the two noise terms correspond 
to an isotropic noise in the complex plane, i.e. $\sigma_1=\sigma_2$.
Therefore, the numerical schemes for both Stratonovich and It\^{o} interpretations can be used to integrate the phases in the case of two noise terms of equal strength.
On the other hand, the two noise terms in \eqref{eq:kurasaka2noise} do not commute.
It has been shown, e.g. in Refs.~\onlinecite{Burrage,BURRAGE1998161,MR2052268}, that the strong order of convergence of all higher order integration methods 
for stochastic differential equations with non-commutative noise cannot be higher than 0.5. Strong order of convergence is defined by the average error made by the time-discretized approximation of the stochastic integration scheme in approximating each individual path of the continuous-time process. 
Therefore we restrict ourselves to using only low-order integration schemes. 
Only in the case of noise in a single direction (i.e. $\sigma_2=0$), high-order methods like the stochastic Runge-Kutta method \cite{Burrage} are used.
In Fig.~\ref{fig:R2} we show the results in the case of two relatively strong noise terms $\sigma_1=\sigma_2= 0.1$~.
The integration is performed with the Euler-Heun scheme for different time steps $h$.
Due to the rotational invariance preserved by two noise terms of equal strength, we have set $\omega=0$.
One can clearly see the formation of two clusters, indicated by values of $R_2$ growing close to one, on a time scale $\sim h^{-1}$. For a weaker noise $\sigma=0.01$, clustering appears much slower. Only initial growth of the second order parameter can be observed at the maximal integration time of $T=4\times 10^5$. 
This dependence of the clustering time scale on the integration step size demonstrates that clustering in this system is a numerical artifact. 
In fact, when we break WS integrability by including a term in the stochastic dynamics \eqref{eq:saka2noise2} proportional to the error in the deterministic integration scheme as in Eq.\eqref{eq:KuramotoEulerError}, i.e.
\begin{eqnarray}
\label{eq:StochKuramotoEulerError}
\dot \varphi_k &=&\omega+\textrm{Im}\left[\left(Z e^{{i}\gamma} + \sigma\xi\right)e^{-i\varphi_k}\right] - \\&&
\frac{\epsilon}{4}\textrm{Im}\left[Z e^{{i}\left(2\gamma-\varphi_k\right)} - Z^* Z_{2}e^{-{i}\varphi_k} - Z^2 e^{{i} 2\left(\gamma-\varphi_k\right)}\right]~,\nonumber 
\end{eqnarray}
we observe robust clustering under dynamics Eq.~\eqref{eq:StochKuramotoEulerError} at a similar time scale as in the original system \eqref{eq:saka2noise2} for an integration time step of $h=\epsilon$ (see Fig.\ref{fig:2timescales}).

\begin{figure}[ht]
 \includegraphics[width=\columnwidth]{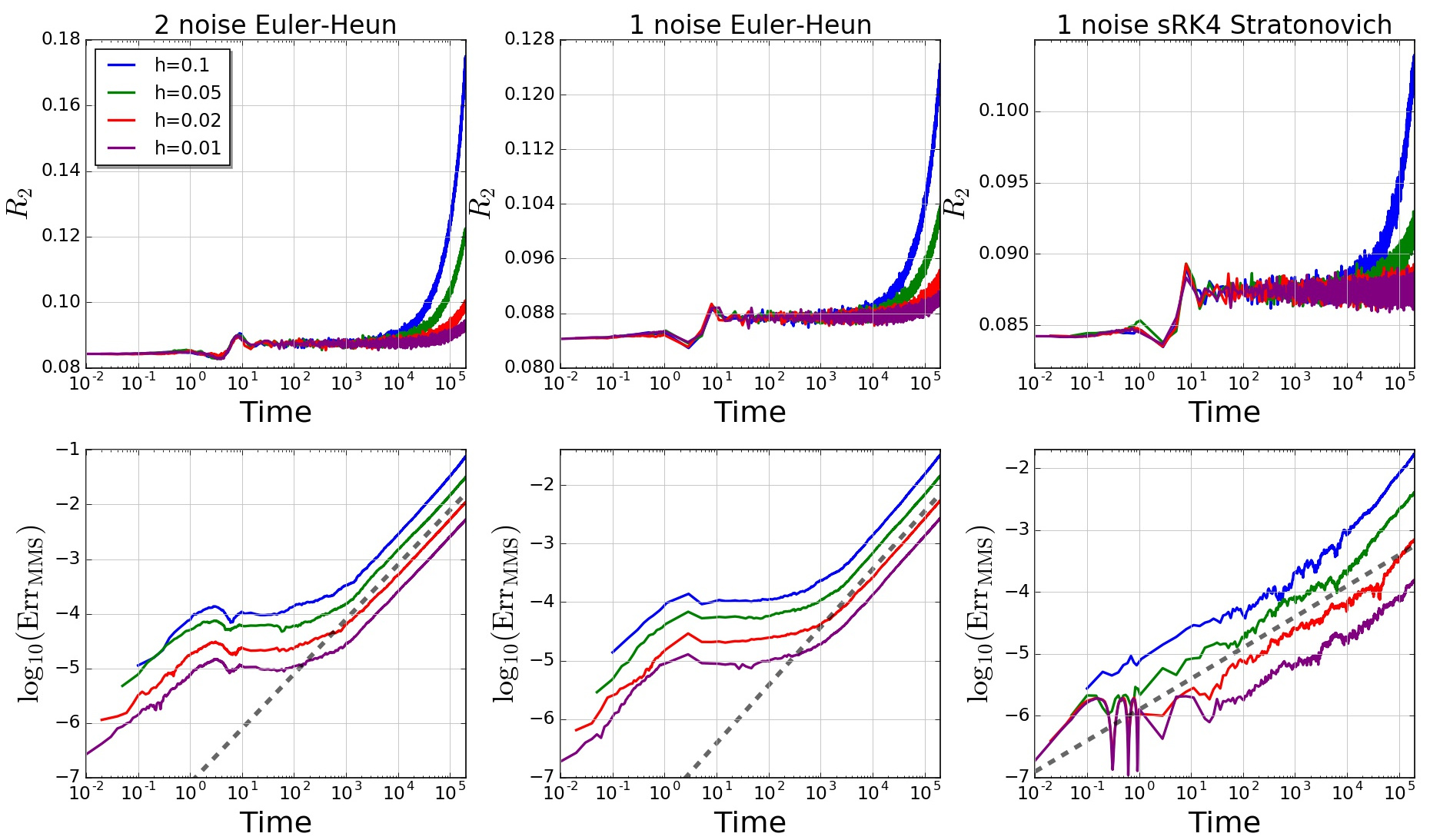}
 \caption{Simulations of Eq.~\eqref{eq:kurasaka2noise} with 1 and 2 noise terms. 
 We show $R_2$ (top panels) and numerical error (bottom panels) 
 as functions of time, for the 
  Euler-Heun and stochastic Runge-Kutta 4-th order methods. Parameters: $N=100$, noise strength $\sigma = 0.01$, 
  phase shift $\gamma=0.6\pi$. 
  Intrinsic frequency $\omega = 0$ for 2 noise terms, 
  and $\omega = 10$ for one noise term. 
  Resulting evolution is averaged over 8 different initial conditions 
 (same for all experiments). 
 The dashed lines in the left and middle columns (Euler-Heun method)
 have slope equal to $1$, whereas in the right column the slope 
 is $0.5$, showing the superiority of the SRK method of integration. }\label{fig:6panel}
\end{figure}

In Fig.~\ref{fig:6panel}, we compare different integration schemes applied to models with one or two noise terms. Here for the cases of two noise terms
(like in Fig.~\ref{fig:R2}) and of one noise (where we set $\omega=10$ because the rotational symmetry is broken), we present results for the 
Euler-Heun scheme, suitable for the Stratonovich interpretation of the stochastic differential equation. Additionally, we show the results
of the stochastic Runge-Kutta scheme (SRK), which is suitable for the one-noise case only, because of the non-commutativity
of the two noise terms mentioned above. One can see that all plots are qualitatively similar, with only some quantitative
differences. As one would expect, the conservation of the constants of motion under the SRK 
scheme is the best, and here also the growth of the second order parameter is rather weak on the
chosen time interval $t<2\cdot 10^5$. We have also performed simulations with the 
Euler-Maruyama scheme with Stratonovich shift for the model \eqref{eq:kurasaka2noise} in the Stratonovich interpretation, both with one noise term and 
with two noise terms (where the Stratonovich shift is zero), all of which yield quantitatively identical results to the Euler-Heun scheme and are therefore not shown.

We can conclude this section by stating that in general, numerical schemes do not conserve
the integrals of motion of the system, and eventually may lead to formation of clusters.
Because the methods for integrating stochastic differential equations have typically lower order than
the deterministic ones, the clusters may be more easily observed in the integration of noisy equations.
In the deterministic case, clustering may not be observed at all if a zero mean field steady state is reached first.
The presence of noise can prolong the time within which the constants of motion 
continue to drift and their deviation from their initial values continues to 
grow until at some point fully synchronized multiclusters are formed. 

As mentioned in section \ref{sec:theory}, the best way to avoid the
numerical artefacts of clustering is to integrate
the Watanabe-Strogatz equations \eqref{eq:WS}, but to accomplish this one has to perform multiple
 M\"obius transforms at each time step for a full time series of the mean field, 
 which may be quite computationally expensive. 
Furthermore, discretization errors are still present in integrating
the low-dimensional dynamics of the M\"obius group parameters. Only multicluster formation would be guaranteed to no longer occur. 

\section{Oscillators with naturally occurring clusters under repulsive coupling 
and common noise - the Van der Pol oscillators} \label{sec:vdp}
Unlike the Kuramoto model, more realistic oscillator models such 
as the Van der Pol oscillator have limit cycles which intrinsically contain 
higher order Fourier terms and additional amplitude dynamics. Under common additive noise and repulsive coupling, 
formation of multiclusters is no longer forbidden and could now naturally occur. 
We consider $N$ identical 
repulsively all-to-all coupled Van der Pol oscillators subject to additive common Gaussian
white noise in one direction
\begin{align}
\dot x_k & = y_i \nonumber \\
\dot y_k & = a (1 - x_k^{2}) y_n - x_n - b \frac{1}{N}\sum_{j=1}^{N}\left(y_j-y_k\right) + \sigma \xi(t)\;.
\label{eq:vdp}
\end{align}
Here $b > 0$ is the repulsive coupling strength, $a$ parametrizes the nonlinearity 
of the Van der Pol oscillators, $\sigma$ is the noise strength, and $\xi(t)$ is a Gaussian white 
noise force. Using phase reduction\cite{winfree} the additive noise term will become multiplicative with the linear phase response function as a factor.

\begin{figure}[ht]
\includegraphics[width=\columnwidth]{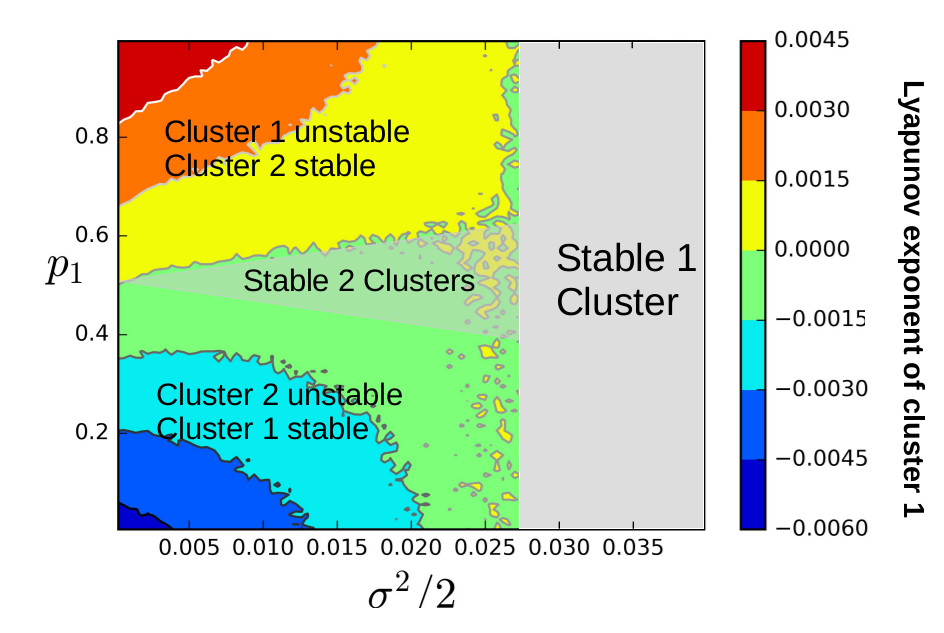}
\caption{Lyapunov exponent diagram for one of the two clusters of repulsively coupled Van der Pol oscillators with additive noise in one direction, similar to diagram \ref{fig:LE}~. 
Contour plot of the Lyapunov exponent of cluster 1 are obtained by numerical integration of the linearized Eqs. \eqref{eq:vdp} for two clusters of relative sizes $p_1$ and $p_2=1-p_1$. System parameter: $a = 1$, $b = 0.01$ correspond to highly nonlinear regime of the Van der Pol oscillator limit cycle. The numerical integration uses the Euler-Maruyama scheme with step size $h = 0.005$. Unlike in Fig. \ref{fig:LE}~, an analytical expression for critical noise strength is hard to obtain. From the simulations we found it to be $\sigma_c^2/2\approx 0.027$. 
The gray region beyond the critical noise strength therefore corresponds to the formation of one cluster under strong noise. Compared to the Kuramoto-Sakaguchi model in Fig.\ref{fig:LE}~, a previously forbidden region of stable 2-cluster appears in the domain $p_1\approx p_2$ below the critical noise strength. As the region with a negative Lyapunov exponent becomes larger as the noise strength increases, it is evident that the common noise stabilizes both clusters.}
\label{fig:stabvdp}
\end{figure}

\begin{figure}[ht]
 \includegraphics[width=0.53\columnwidth]{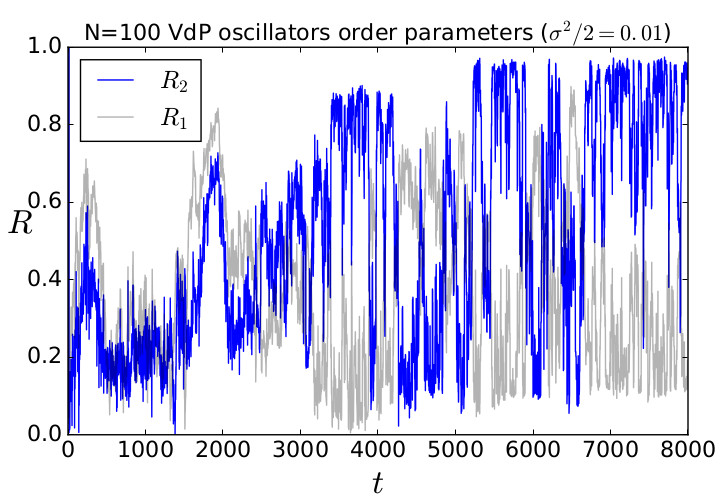}
 \includegraphics[width=0.45\columnwidth]{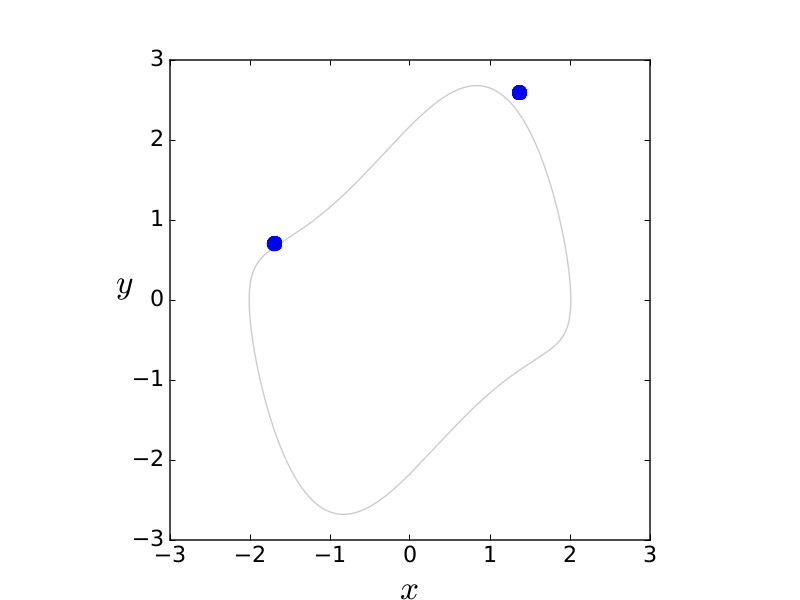}
 \caption{Direct simulation of Van der Pol oscillator ensemble of size $N=100$ under weak common additive noise and repulsive coupling results in stable two clusters with relative sizes $p_1=53\%$ and $p_2=47\%$ after a transient. Left: time series for order parameters $R_1$ and $R_2$ during the initial transient from normal Gaussian random initial conditions in the $(x,y)$ plane. Right: snap shot of two stable clusters formed after integration time $T = 33000$. System parameters are $a=1.0$, $b=0.01$ and $\sigma^2/2 = 0.01$. Euler-Maruyama integration scheme with $h = 0.001$ is used. This is consistent with the negative evaporation Lyapunov exponents for both clusters within the triangular parameter region in Fig.~\ref{fig:stabvdp}. }\label{fig:directsimvdp}
\end{figure}

With a similar approach to that of section~\ref{sec:linstab}, one can determine the Lyapunov exponents for the two-cluster state
in~\eqref{eq:vdp} numerically by integrating a perturbation from one of the clusters in the linearized dynamics of the two cluster system.
Contrary to the case of the Kuramoto model, presented in Fig.~\ref{fig:LE},
now in Fig.~\ref{fig:stabvdp} we see that the two-cluster state with $p_1\approx p_2$ is 
locally stable, which is confirmed in Fig.~\ref{fig:directsimvdp} by direct
simulations of Eq.\eqref{eq:vdp}.  Here, we defined the Kuramoto
order parameters using the phases defined by virtue of Poincar\'e sections. One oscillator
has been chosen as a reference, and the moments of time $\overline{t}_1,\overline{t}_2,\ldots$
at which it crosses half-line $(x>0,y=0)$ have been determined. Then the phase differences
of all other oscillators to the reference one at time $\overline{t}_m$ were
defined as $2\pi (t_m^{(k)}-\overline{t}_m)/(\overline{t}_{m+1}-\overline{t}_m)$. Here
$t_m^{(k)}-\overline{t}_m$ is the time needed for an oscillator
 with index $k$  to reach the Poincar\'e section from its position at time
 $\overline{t}_m$.

In general, clustering strongly depends on the level of nonlinearity, described by $a$. For large $a$ ($a=1$, $b=0.01$),
in the deterministic case three clusters can be observed. In the presence of noise, the picture is not so clear as
several different cluster states may appear depending on the realization of initial conditions and of the noise, 
but a tendency at least to temporal formation of clusters is clearly observed. For small
values of nonlinearity parameter $a$, typically non-clustered states are observed both with and without noise. This is to be expected, since the
Van der Pol oscillator with a weakly anharmonic limit cycle has comparably much smaller higher order Fourier terms in its phase response function.
In general, dynamic complexity of systems like~\eqref{eq:vdp} with non negligible amplitude dynamics can be very high, with chimera-like states becoming possible (i.e. where 
clusters coexist with dispersed elements), and a full characterization is beyond the scope of this paper.

From the above observation we can therefore conclude that there exists a qualitative difference between the dynamics of the phase oscillator model (e.g. Fig. \ref{fig:LE}), and that of the more general oscillator model with additional amplitude dynamics (e.g. Fig. \ref{fig:stabvdp}), specifically under a repulsive coupling and common noise: while under Kuramoto-Sakaguchi model clusters are not allowed to form, under Van der Pol model they are naturally forming and are stablized by the common stochastic forces. 

\section{Conclusions}
In this study, we apply the Watanabe-Strogatz (WS) theory \cite{ws94} to the Kuramoto-Sakaguchi 
model of repulsively coupled phase oscillators under common noise,
studied previously in Ref.~\onlinecite{Mikhailov}~. Our main result is that although both the WS theory and the stability
analysis of clustered states exclude appearance of clusters as observed in~\onlinecite{Mikhailov},
these observations can be generally explained as artefacts from the finite accuracy of numerical simulations. The correct long term behavior for repulsively coupled phase oscillators under common noise is either an incoherent state with no clustering (when the common noise has weaker effect compared to the repulsive coupling) or a completely coherent state (when the common noise has a stronger effect compared to repulsive coupling).
We study the numerical errors of different deterministic and stochastic schemes by monitoring the evolution of the constants of motion
which are conserved under the exact dynamics. It should be stressed that the conclusions of WS theory
only apply to a restricted class of phase oscillators which approximate weakly coupled, weakly nonlinear limit cycle oscillators. Violation of 
WS integrability occurs naturally in general coupled oscillator systems.
We show that in the case of repulsively coupled Van der Pol oscillators noise-induced or deterministic clustering can indeed be easily observed in regimes
of larger nonlinearity.
Due to the limitation of the Kuramoto-Sakaguchi system in describing real-world oscillator models or even more complicated coupled systems of differential equations, in terms of numerics, this paper presents only a cautionary tale. For most types of high dimensional coupled differential equations, a hidden low-dimensional dynamics such as present in Kuramoto-type system is not available, nor do there often exist integrals of motion. For these systems, often the only way to measure or to gauge numerical errors is by using integration steps which are as small as possible, and to compare results under various degrees of numerical accuracies.

\section*{Acknowledgments}
This paper is developed within the scope of the IRTG 1740/TRP 2015/50122-0, funded 
by the DFG/ FAPESP. We thank Denis S. Goldobin and Michael Zaks for valuable discussions. 
C.Z. also acknowledges the financial support from China Scholarship Council (CSC).
Work of A.P. is supported by Russian Science Foundation (Grant Nr.\ 17-12-01534).

\bibliographystyle{unsrt}
\bibliography{biblio}

\end{document}